\documentclass[12pt]{article}
\usepackage{epsfig,subfigure,rotating,a4}
\begin{document}
\title{Investigation of A Lattice Boltzmann Model with a Variable Speed of Sound}

\author{J M Buick$^1$, J A Cosgrove$^2$\\
$^1$ Physics and Electronics, University of New England, Armidale\\
NSW 2351, Australia\\
$^2$ School of Physics, The University of Edinburgh\\ 
Edinburgh EH9 3JZ, Scotland, UK}

\maketitle
\begin{abstract}
A Lattice Boltzmann model is considered in which the speed of sound can be varied
independently of the other parameters. The range over which the speed of 
sound can be varied is investigated and good agreement is found between 
simulations and theory. The onset of non-linear effects due to variations
in the speed of sound is also investigated and good agreement is again found
with theory.
It is also shown that the fluid viscosity is not altered
by changing the speed of sound. 
\end{abstract}

\section{Introduction}
The lattice Boltzmann model (LBM) 
is a numerical technique for fluid simulation 
which has become increasingly popular in recent years \cite{Suc2b,Chen1:art,
Succi1:bk,wol1b}. 
The LBM originates from the lattice gas model (LGM) 
\cite{har1, har2, fri1, fri2}
were fluid particles are constrained to move on a regular lattice such that
their collisions conserve mass and momentum. The particles are further 
constrained 
to move with unit velocity and with a maximum occupancy of
one particle per grid direction per grid site. The evolution of the
LBM from the lattice gas model involved a number of key developments.
The Boolean particle number (1 if a particle is present and 0 otherwise)
was replaced by a real number, later recognized as 
the distribution function, representing an ensemble average of the particle
occupation \cite{mcn1b}; this removed the noise associated with the relatively
small number of fluid particles represented in the LGM. Obtaining
an ensemble average of the particle collisions is not straightforward but
the process was simplified to depend only on the link 
direction \cite{hig1b} and the isotropy of the model \cite{hig2b}. Finally the
particle-particle collisions were replaced by considering the distribution
functions relaxing toward a Maxwell-Boltzmann equilibrium distribution 
\cite{che1b}.\\

The LBM
is referred to as
an incompressible technique because the LBM scheme can be shown to satisfy
the incompressible Navier-Stokes equation in the limit that the density,
$\rho$, does not vary in space or time. However, when applying the LBM, there
is no restriction on the density to remain constant and in many applications
the density will vary in space and/or time. In practice the `incompressible'
nature of the LBM is interpreted as requiring that any density variations are
small or that the density varies only slowly in space and time. In this limit
it is possible to simulate phenomena such as acoustic waves where a density
variation is required, provided the variation is small \cite{buickac1,buickac2,
haydock1}.\\

In the LBM,
pressure is defined through the equation of state,
see for example \cite{Chen1:art}, $p = c^2 \rho$, where
$c$ is the speed of sound which takes 
a fixed value in any simulation, dependent
only on external factors such as the shape of the simulation grid. 
Thus we see that the
speed of sound controls the relationship between the density and the
pressure and therefore the compressibility of the fluid. To model fluids
with different compressibility we require to be able to change the
speed of sound in the simulation. Here we consider a model 
with a variable speed of sound and investigate
the non-linear aspects of sound wave propagation.\\

\section{The lattice Boltzmann model}
In this section we briefly consider the standard LBM on a square grid in
two dimensions.
In the LBM the distribution functions, $f_i(\mbox{\boldmath{$x$}},t)$,
for $i$= 0-8 evolve on 
a regular grid according to the Boltzmann equation \cite{qia1b}, as
\begin{equation}
f_i(\mbox{\boldmath{$x$}}+\mbox{\boldmath{$e$}}_i,t+1) - 
f_i(\mbox{\boldmath{$x$}},t) = - \frac{1}{\tau}
(f_i - \overline{f}_i),
\label{eqn:bolt}
\end{equation}
where
$\mbox{\boldmath{$e$}}_i$ = $(\cos[\pi(i-1)/2],\sin[\pi
(i-1)/2])$ for $i$ = 1-4, and
$\mbox{\boldmath{$e$}}_i$ = $\sqrt{2}(\cos[\pi(i-9/2)/2],\sin[\pi
(i-9/2)/2])$ for $i$ = 5-8 are link vectors on the grid and 
$\mbox{\boldmath{$e$}}_0$ = (0,0). 
The left hand side of equation (\ref{eqn:bolt}) represents streaming of the
distribution function from one site to a neighbouring site. The
right hand side is the Bhatnagar, Gross and Krook (BGK) collision
operator \cite{bha1b,che1b,che3b}.
The equilibrium distribution function
$\overline{f}_i$ is given by \cite{qia1b}
\begin{equation}
\overline{f}_i = w_i \rho \left[ 1 + 3\mbox{\boldmath{$e$}}_i
\cdot \mbox{\boldmath{$u$}} + \frac{9}{2} (\mbox{\boldmath{$e$}}_i
\cdot \mbox{\boldmath{$u$}})^2 - \frac{3}{2}
u^2 \right],
\end{equation}
where the fluid density, $\rho$, and velocity, $\mbox{\boldmath{$u$}}$
are found from the distribution function as
\begin{equation}
\rho = \sum_{i=0}^{8} f_i \quad \mbox{and} \quad
\rho \mbox{\boldmath{$u$}} = \sum_{i=0}^{8} f_i \mbox{\boldmath{$e$}}_i
\end{equation}
and where $w_0 = 4/9$, $w_i$ = 1/9 for $i$=1-4, and $w_i$ = 1/36
for $i$ = 5-8. The relaxation time, $\tau$, determines the rate at which the
distribution functions relax to their equilibrium values; it
determines the fluid viscosity, $\nu$, as
\begin{equation}
\nu = \frac{(2 \tau -1)}{6}.
\end{equation}

Equation~(\ref{eqn:bolt}) can be modified by the addition of an extra source
term. This approach has been used to implement a body force \cite{buickg1}
and also to simulate non-ideal equations of state such as phase separation
\cite{sha1b}.

\section{A lattice Boltzmann Model with a Variable Speed of Sound}
One method for achieving a variable speed of sound was proposed by
Alexander {\em et al.} \cite{Alexander1:art} who considered a 
model on a hexagonal grid with an altered equilibrium distribution function
in which the ratio of `rest particles' ($f_0$) to `moving particles' ($f_i,
i \ne 0$)
can be altered. Simulations performed using this model
\cite{Alexander1:art} showed that the speed of sound can be
varied between 0 and approximately 0.65 with good agreement
between theory and simulation for speed less than 
approximately 0.4. Above 0.4 there is some deviation between 
theory and simulation. In this model the viscosity was also
found to be a function of the variable speed of sound.\\

An alternative approach was proposed by Yu and Zhao \cite{Yu:art}.
They introduced an attractive force which produced a variable speed of sound
which was dependent on the amplitude of the introduced force. The model was
verified by measuring the Doppler shift and the Mach cone for Mach numbers 
less than and greater then unity respectively.\\

Here we consider the following
LBM:
\begin{equation}
f_i(\mbox{\boldmath{$x$}}+\mbox{\boldmath{$e$}}_i,t+1) - 
f_i(\mbox{\boldmath{$x$}},t) = - \frac{1}{\tau}
(f_i - \overline{f}_i) + 3 w_i \alpha \mbox{\boldmath{$\nabla$}} \rho \cdot 
\mbox{\boldmath{$e$}}_i,
\label{eqn:nbm}
\end{equation}
where
$\overline{f}_i$ and $w_i$ are defined as previously and the additional term
on the right-hand side of equation~(\ref{eqn:nbm}) represents
a body force $\alpha \nabla \rho$, see for example \cite{buickg1}.
This is effectively the same as the model of Yu and Zhao \cite{Yu:art},
except that the forcing term is expressed explicitly as proportional
to the density gradient and the amplitude term, $\alpha$ is not restricted to
be positive.
Performing a Chapman-Enskog expansion \cite{fri2} 
we can write the Navier-Stokes equation
in the form
\begin{equation}
 \partial_t u_{\alpha} + u_{\beta} \partial_{ \beta}  u_{\alpha}
= -\frac{1}{\rho} \partial_{ \alpha} \left( c_1^2 - \alpha \right) \rho
+ 
\nu \partial_{\beta} ( \partial_{\alpha} u_{\beta} + \partial_{\beta} u_{\alpha}),
\label{forgns}
\end{equation}
where 
\begin{equation}
\nu = \frac{2\tau - 1}{6}
\label{eqn:visc}
\end{equation}
as before, and we have assumed that the derivatives of $\rho$
can be neglected except where they appear in the pressure term.
This is the Navier-Stokes equation for a fluid which has pressure
$p=c_e^2 \rho$, where $c_e$ is the effective speed of sound which is given by
\begin{equation}
c_e = \sqrt{\left( c_1^2 - \alpha \right)}.
\label{cef}
\end{equation}
Thus by introducing an additional term to the Boltzmann equation
which acts as a body force proportional to the density gradient,
we have included a force which increases or decreases (depending on the
sign of $\alpha$) the pressure forcing term in the Navier-Stokes equation,
and hence the speed of sound in the model.\\

A number of models, based on the cellular automata (CA) approach
of the LGM, have also been applied to simulate acoustic
waves as well as Burgers' equation. Mora \cite{Mora1:art} introduced the
phononic lattice solid model which obeyed a Boltzmann equation and 
satisfied the acoustic wave equation in the
macroscopic limit. The Boltzmann equation was solved using a finite-difference 
scheme. This model was further developed \cite{Mora2:art} in the phononic 
lattice solid by interpolation model in which the particle number densities
move along the lattice links in the same manner as the distribution functions
in the LGM and LBM. Unlike the lattice gas particles they travel at
different speeds and so, in general, do not arrive at a grid site after each
time step. The number densities are therefore interpolated to find the value 
at each site.

CA models for Burgers' equation have also been considered. 
The model of Boghosian and Levermore \cite{Boghosian1:art} was based on the LGM
approach and was shown to follow the solution of Burgers' equation. The
convergence of the model was also established \cite{Lebowitz1:art}. Following 
the development of the LBM from the LGM \cite{mcn1b, hig1b, hig2b},
Elton \cite{Elton1:art} considered a CA model for Burgers' equation based
on the mean occupation number rather than discrete particles. The model 
compared 
favourably with a finite difference solution of Burgers' equation. A quantum
lattice gas model \cite{Yepez1:art} and an intrinsically stable entropic model
\cite{Boghosian2:art} have also been proposed recently. Each of these models
provides a simulation method which satisfies Burgers' equation. This is 
different to the model investigated here which satisfies the Navier-Stokes
equation and is also capable of simulating non-linear acoustic waves whose
behaviour, for the case $\alpha = 0$, has been shown to be in good agreement
with Burgers' equation \cite{buickac2}. This means that it is capable of
simulating the interaction between acoustic waves and fluid flow,
see for example \cite{haydock1, haydock2, Buickradforce:Art, haydock3}.\\

\section{Numerical Simulations}

\begin{figure}[p]
\centering\mbox{\epsfig{file=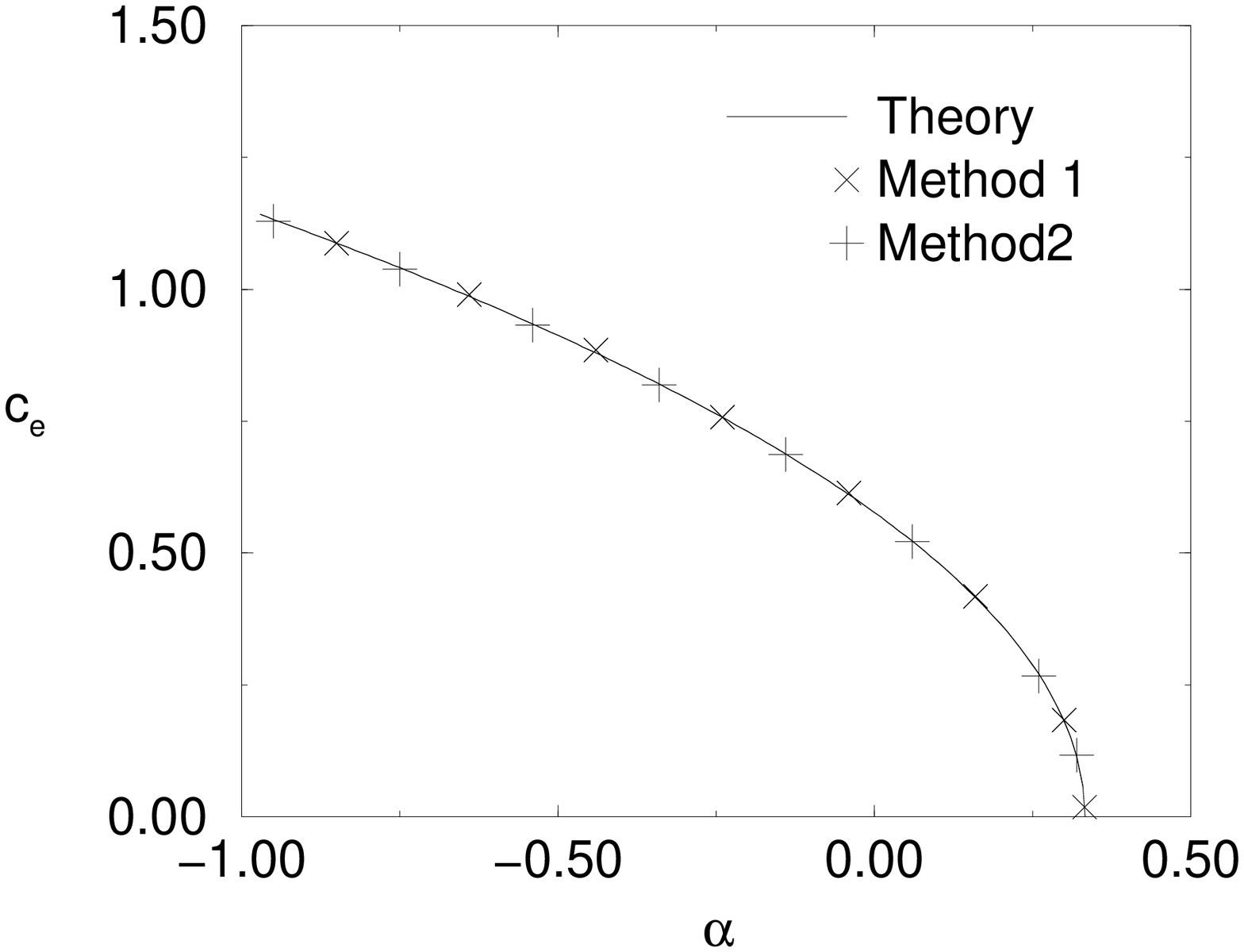,width=0.78\linewidth}}
\caption{The effective speed of sound, $c_e$ as a function of 
the forcing 
magnitude $\alpha$.}
\label{fig:fullres}
\end{figure}

The variation in the speed of sound was investigated in the following
simulations of plane acoustic wave propagating in an unbound
media. This if effectively a one dimensional problem so a
gird was used with only 4 points in the $y$-direction. The grid
had $L$ points in the $x$-direction and
was initialised with zero velocity and unit density. Periodic
boundary conditions were applied at $y$ = 0 and $y$ = 4. At $x$ = 0 a boundary
condition was applied in which the density was varied in a sinusoidal
manner with a
period of 500 time-steps, the velocity at this grid boundary 
was maintained at zero. At $y$ = $L$ a boundary condition of
$\rho$ = 1 and $\mbox{\boldmath{$u$}}$ = 0 was applied.
$L$ was selected to
be large enough so that the density disturbance does not reach $x$ = $L$ during
the measurement window.
From these simulations the speed of sound was found in two ways:
1) from the time taken for
the wave to pass between two points at a known separation; and 2) 
since we are dealing with a low amplitude plane harmonic wave,
$c_e$ can be derived from \cite{Kinsler1}
$u(\mbox{\boldmath{$x$}}^*,t^*)/c_e = 
[ \rho(\mbox{\boldmath{$x$}}^*,t^*)- \rho_0] / \rho_0$, 
where $u(\mbox{\boldmath{$x$}}^*,t^*)$ and $\rho(\mbox{\boldmath{$x$}}^*,t^*)$ 
are the fluid velocity
and density respectively, at some position $\mbox{\boldmath{$x$}}^*$ and 
time $t^*$; here we selected $\mbox{\boldmath{$x$}}^*$ and $t^*$ such that
they correspond to a pressure and velocity maximum. 
The results are shown in figure \ref{fig:fullres} where there is excellent
agreement between the measured values of $c_e$ and the theoretical values
given by equation~(\ref{cef}) over a range of $\alpha$. At $\alpha =
1/3$ we have  $c_e = 0$ giving an upper limit for $\alpha$. Now 
$\alpha = -2/3$ gives $c_e = 1$ which would appear to be an upper limit
since this is the speed with which the distribution functions propagate and
so is the maximum speed information about the density, pressure and velocity
can be transmitted through the fluid. Initial simulations were performed
using a simple forward difference scheme to calculate the pressure gradient.
In this case the model became unstable as $c_e$ approached unit, as would be
expected. When a central difference technique was applied to calculate the
density gradient the model was found to remain stable as $c_e$ approaches
unity. Indeed it was found to remain stable for higher values of $c_e$ 
($\alpha <$ -2/3) and the simulated density waves were observed to 
propagate with $c_e >$ 1. The transfer of information at a speed greater than 
unity can be understood with the introduction of a central difference
scheme to calculate the `pressure' forcing term. Consider a 
density distribution where $\rho(x) = \rho_0$ everywhere at t=0, except for
$\rho(x_0) = \rho_0 + \delta \rho$. Then at $x=x_0+1$ the density gradient
(calculated using a central difference scheme)
will be non-zero. In the case of $\alpha <$ 0 this will give a forcing term
that will cause the value of the distribution function travelling in the 
positive direction at $x_0+1$ to increase. This information (about a density
increase at $x_0$) will then be passed on to $x_0$+2 at t=1 by the streaming
action. As seen in figure \ref{fig:fullres} this is a limited effect; however,
it suggests that LB schemes in which the speed of sound is significantly 
greater than unit may be possible if a longer range action is applied.
The largest value of $c_e$ which was observed to remain stable was $c_e$ = 
1.125. 
The validity of simulations with $c_e >$ 1 has not been established and
requires further investigation. The remainder of the simulations presented here
have have $c_e <$ 1.\\

To further investigate the effectiveness of the model two progressive
sound waves with $\lambda = 4000$ were
initialised with the same velocity amplitude: $u(x,t=0) = u_{0}
\sin(kx)$, $v(x,t=0)=0$ and density profile
$\rho^{(i)}(x,t=0) = 
\rho_{0}^{(i)}(1 + u^{0}/c_{e}^{(i)})\sin(kx)$,
for $i$ = 1, 2; where $c_{e}^{(i)}$ is found from $\alpha^{(i)}$
with $\alpha^{(1)}$ = 0.2933 and $\alpha^{(2)}$ = -0.47667. 
This
gives $c_e^{(1)}$ = 0.2 and $c_e^{(2)}$ = 0.9 such that
$c_e^{(2)}/c_e^{(1)} = 4.5$.
The ambient density $\rho_{0}^{(i)}$ is also free to
be varied, however, in the simulations we used
$\rho_{0}^{(1)} =\rho_{0}^{(2)} = 1$ and $u_0 = 0.002$. 
For these two waves we can calculate the shock 
development distance given by
\begin{equation}
\sigma^{(i)} = \frac{1}{\epsilon M^{(i)}  k},
\label{eqn:sfd}
\end{equation}
where 
$M^{(i)} = u/c_e^{(i)}$ is the Mach number, $k= 2 \pi/ \lambda$
is the wavenumber and $\epsilon = 1$. This is the distance over
which an initially sinusoidal wave in an inviscid fluid
will develop into a discontinuous wave with a sawtooth shape.
Here we have $\sigma^{(1)} \simeq 16 \lambda$ and
$\sigma^{(2)} \simeq 72 \lambda$ so we expect wave (1) (in
the fluid with the lower speed of sound) to exhibit stronger
non linear effects than wave (2).\\

\begin{figure}[p]
\centering\mbox{\epsfig{file=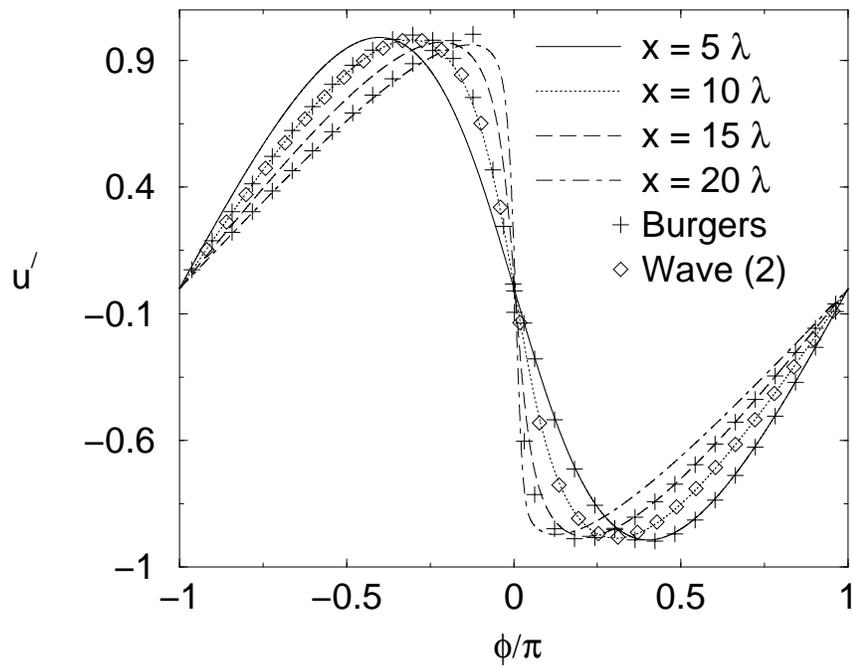,width=0.78\linewidth}}
\caption{The nonlinear development of the sound wave for $c_e$ = 0.2 over the
first twenty wavelengths. Also shown for comparison is the numerical solution
of Burgers' equation (+) and the LBM simulation of a sound wave with
$c_e = 0.9$ at $x = 45 \lambda$ (diamonds).}
\label{fig:non}
\end{figure}

\begin{figure}[p]
\centering\mbox{\epsfig{file=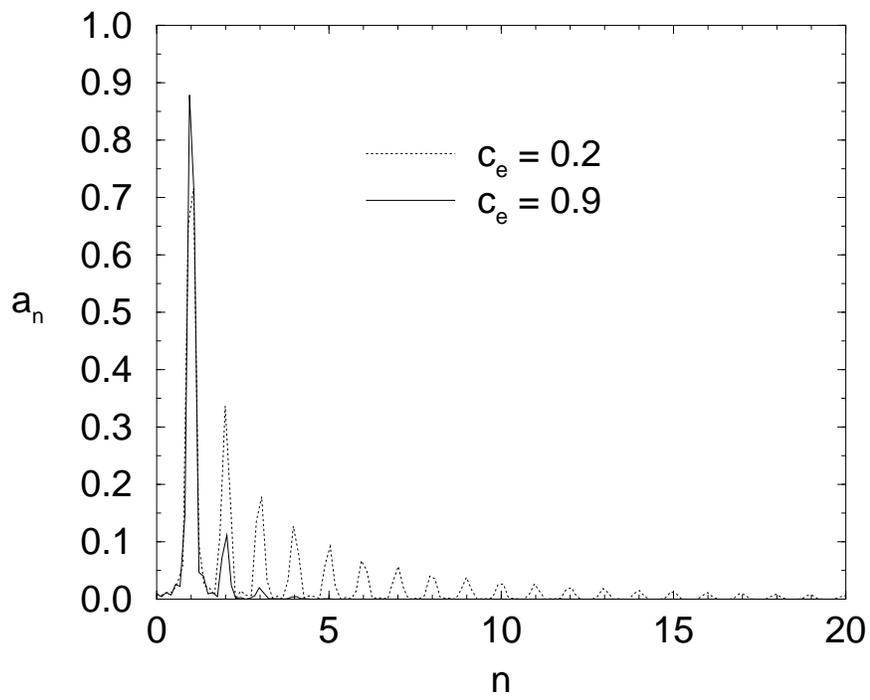,width=0.78\linewidth}}
\caption{Amplitude of the spectral components for wave (1) and wave (2) 
calculated over just over six wavelengths starting at $x=14 \lambda$.}
\label{fft}
\end{figure}

Following \cite{buickac2} periodic boundary conditions were
applied on all grid boundaries and the waves were allowed to 
propagate. Figure \ref{fig:non} shows the normalised
velocities $u' = u/u_0$; plotted as a function of $\phi = \omega(t-x/c_e)$
after the waves has propagated distances of $5 \lambda$,
$10 \lambda$ and $15 \lambda$. Here $\omega$ is the angular frequency of the
acoustic wave. It can clearly be seen that 
wave (1) exhibits strong non-linear behaviour over the first twenty 
wavelengths. Also shown in figure~\ref{fig:non} is the
numerical solution of Burgers' equation \cite{buickac2} which
describes the wave form in a viscous fluid in terms of a truncated sum of
$N$ harmonics. Good agreement is observed for $x= 5, 10$ and
15 $\lambda$. For $x= 20 \lambda$ a suitable value of $N$ could not be found
to give a smooth profile. Even at large $N$ oscillations were observed on the
wave form around $\phi = 0$ and $\phi= \pm 1$. For this reason the solution
of Burgers' equation is only shown away from these points for $x=20$ where there
is again good agreement with the Boltzmann simulations.
The LBM simulation of wave (2) at $x = 45 \lambda ( = 10 (c_e^{(2)}
/c_e^{(1)}) \lambda)$ is also shown
in figure~\ref{fig:non} for comparison. The wave form is similar to that for wave (1) at $x = 10 \lambda$ as would be expected from equation~(\ref{eqn:sfd})
since both waves were run with a relatively low viscosity ($\nu =
0.01$).
A comparison with wave (1) is given in figure \ref{fft} which shows
the amplitudes, $a_n$, of the fundamental ($n=1$) and higher
($n > 1$) harmonics. The results in figure \ref{fft} were
obtained by performing a fast Fourier transform on the velocity
between $x= 14 \lambda$ and $x > 20 \lambda$ where the smallest number of
point which were an exact multiple of 2 were used.\\

\begin{figure}[p]
\centering\mbox{\epsfig{file=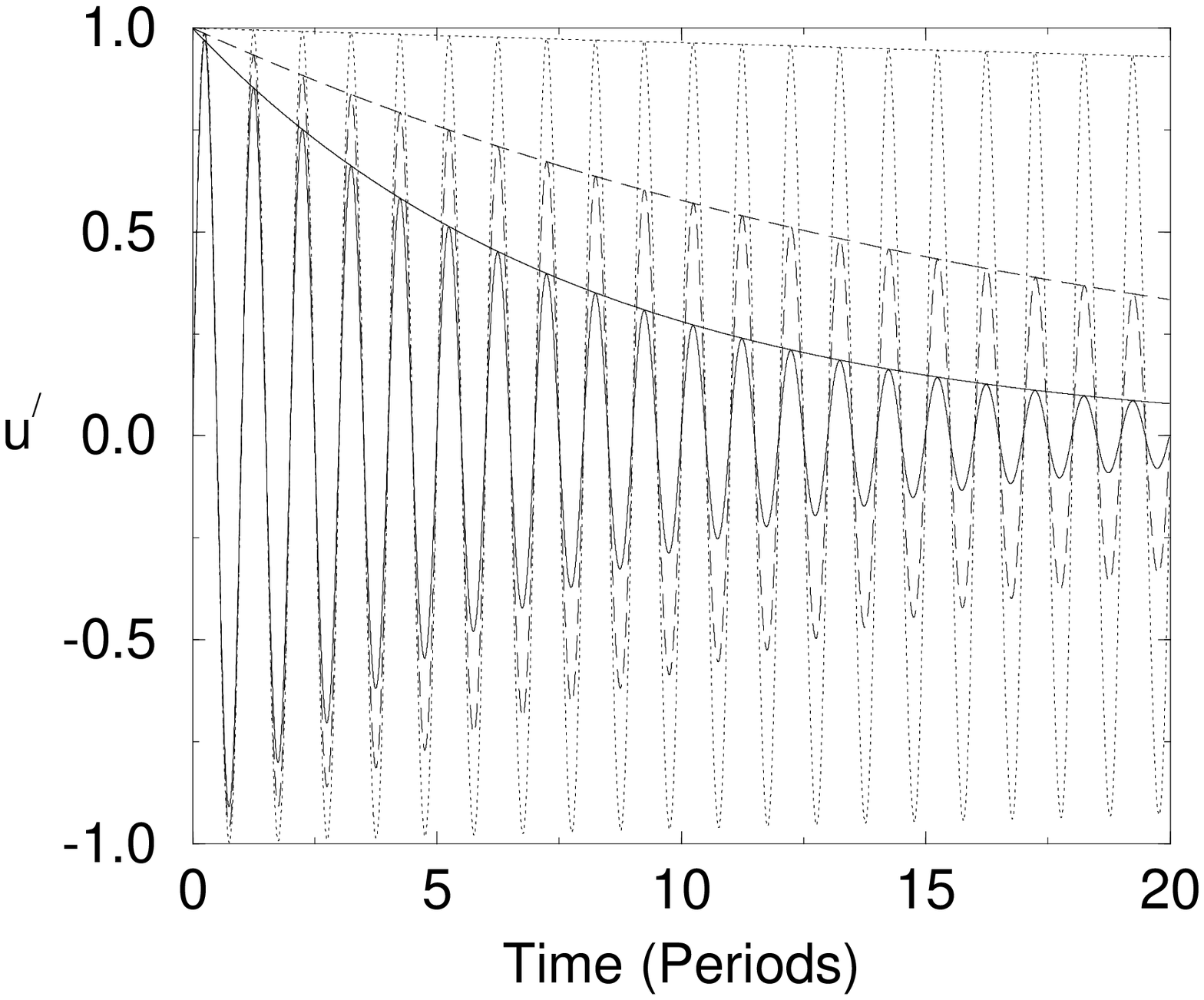,width=0.78\linewidth}}
\caption{The propagation of wave(2) for three selected values of $\tau$. Also 
shown is the exponential decay rate predicted by linear theory.}
\label{visc}
\end{figure}

Figure \ref{visc} shows the long term behaviour of wave (2)
for three different fluid viscosities given by $\nu$ = 1/30, 1/2 and 7/6
,
the value of $\lambda$ was 400.
Relatively large values of $\tau$ were selected to ensure that the wave was
damped rapidly.
The results are in good agreement with the exponential decay rate
predicted by linear theory, see \cite{Lord,Kinsler1,buickac1}, which is also
shown.\\ 

\begin{figure}[p]
\centering\mbox{\epsfig{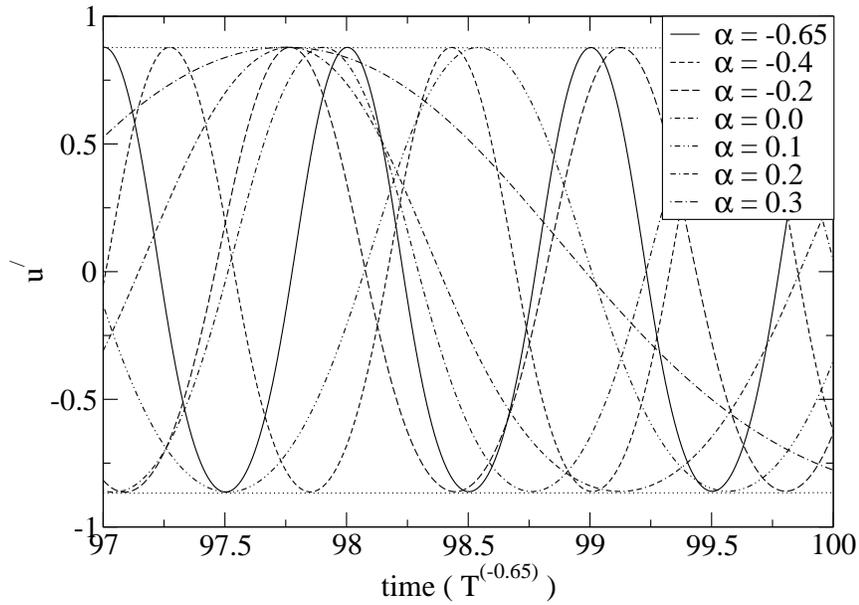}}
\caption{Propagation of wave(2) for $\tau$ = 0.52 and selected values of 
$\alpha$ for $97T^{(-0.65)} \le t \le 100T^{(-0.65)}$, where $T^{(-0.65)}$ 
is the period of the wave when $\alpha = -0.65$.}
\label{allalpha}
\end{figure}

\begin{figure}[p]
\centering\mbox{\epsfig{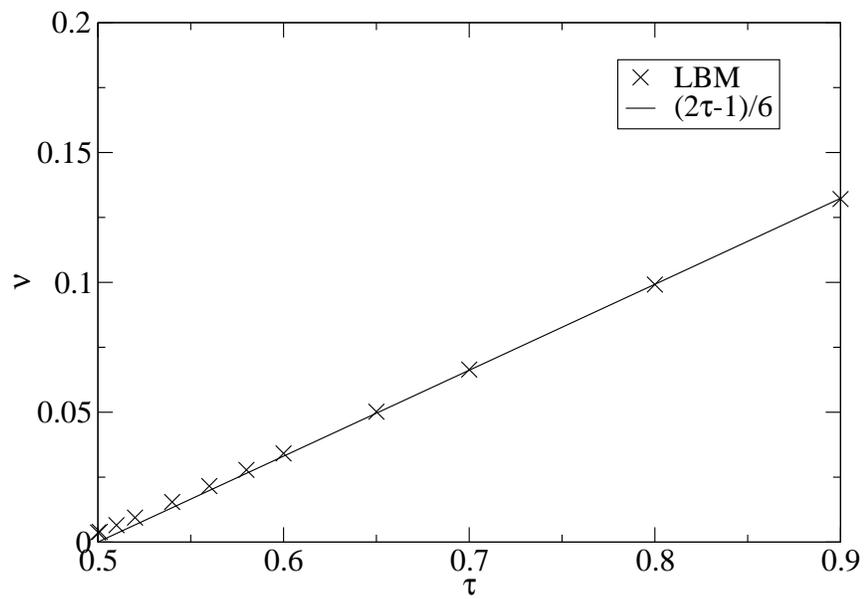}}
\caption{The fluid viscosity calculated from the decay of a wave after 100 periods with $\alpha$
= -0.6, as a function of the relaxation parameter. The solid line 
represents the theoretical expression.}
\label{decay}
\end{figure}

The long term behaviour of wave (2) was also considered in fluids with
a lower viscosity. The decay rate per unit time is independent of the
speed of sound and depends only on the wavelength of the wave and the
viscosity of the fluid. This is shown in figure \ref{allalpha} which depicts
wave (2) for a range of $\alpha$ values after around 100 periods for the 
wave with 
$\alpha = -0.65$. It is clear from figure \ref{allalpha} that the damping rate
is the same for each value of $\alpha$; including $\alpha$ = 0 which corresponds
to the standard LBM with no additional body force. The viscosity of the fluid
was obtained from the simulations as 
\begin{equation}
\nu = -\frac{1}{k^2t}\ln\left(\frac{u_{t=100T}}{u_{t=0}}\right)
\end{equation}
and is shown in figure \ref{decay} for $\alpha = -0.6$ for values of $\tau$
between 0.501 and 0.9. Figure \ref{decay} shows good 
agreement between the
measure viscosity and the theoretical value at large viscosities.
At lower viscosities their is some deviation. For $\alpha$ = -0.6 the largest 
difference
observed between the simulated wave amplitude after 100 periods and the
theoretical value was less then 4 \%. It is well known that the LBM becomes
unstable as $\tau \rightarrow$ 0.5 with noise being introduced into the 
simulation which eventually becomes unstable. This has been observed elsewhere,
for example \cite{Wilde:ip} where a filter was introduced to reduce the noise. 
Here a filter was not required
and the introduction of the body force term was not observed to 
significant alter the onset of instabilities.
The good agreement between 
the simulations and the theory indicates that
the introduction
of a variable speed of sound using the method proposed here
does not effect the viscosity of the fluid, as predicted
by equations~(\ref{forgns}) and (\ref{eqn:visc}).\\

\section{Conclusions}

A new lattice Boltzmann model has been devised in which the
speed of sound can be varied. This has been demonstrated
in a number of simulations in which the speed at which a disturbance
propagates in a fluid was found to agree with the theoretical
speed of sound. Further, it has also been shown that the
non-linearity of a pressure wave depends not only on its amplitude
but also on the speed of sound in the medium through which it
is propagating, as would be expected, and good agreement was found with the
numerical solution of Burgers' equation.
Compared to a previous model \cite{Alexander1:art}, the present model has a
larger range over which the speed of sound can be varied and shows better 
agreement  with theory over the full range.
Additionally the method by which the
variable speed of sound is introduced does not change the
viscosity of the fluid.
The speed of sound is varied in the new model by
a free parameter, $\alpha$, which was kept constant in space and time
during the simulations presented here. This is not, however,
a fundamental requirement and it is envisaged that this model
could be applied such that the speed of sound varies within
the simulation. For example, in a simulation of a liquid-gas or 
a binary fluid mixture, the speed of sound could be varied
as a function of the fluid species.
This could be done in much the same was as a species dependent 
relaxation time has been used to simulate binary fluids
with different viscosities, see for example \cite{Langaas1:art}.
Alternatively, in a single phase fluid simulation, the
speed of sound could be varied in a pre-determined manner to
account for an external influence; for example a temperature
gradient or an acoustical lens. 
The model can also be applied to simulate fluids
with different compressabilities which extends the range of
application of the LBM.

\bibliography{paper,1,bolt,ca,new_add,new,template}

\end{document}